\documentclass[aps,prx,superscriptaddress,notitlepage, amsfonts,longbibliography, twocolumn]{revtex4-2}

\usepackage[dvipdfmx]{graphicx}

\usepackage{amsmath,amssymb,amsthm,mathtools,mathrsfs}
\usepackage{srcltx}

\usepackage{bm,braket,comment}
\usepackage{amsmath, amssymb,amsthm,mathtools}
\usepackage{xcolor}
\usepackage{longtable}
\usepackage{enumerate}
\usepackage{ulem}
\usepackage{multirow}
\usepackage{here}
\usepackage{comment}
\usepackage{url}

\newtheorem{theorem}{Theorem}
\newtheorem{definition}{Definition}

\newcommand{\bout}[1]

\usepackage{hyperref}
\hypersetup{colorlinks=true,linkcolor=blue,citecolor=blue,urlcolor=blue}

\allowdisplaybreaks[4]
\begin{document}

\title{Unitary Designs of Symmetric Local Random Circuits}

\author{Yosuke Mitsuhashi}
\email{mitsuhashi@noneq.t.u-tokyo.ac.jp}
\affiliation{Department of Basic Science, University of Tokyo, 3-8-1 Komaba, Meguro-ku, Tokyo 153-8902, Japan}

\author{Ryotaro Suzuki}
\email{ryotaro.suzuki@fu-berlin.de}
\affiliation{Dahlem Center for Complex Quantum Systems, Freie Universität Berlin, Berlin 14195, Germany}

\author{Tomohiro Soejima}
\email{tomohiro\_soejima@g.harvard.edu}
\affiliation{Department of Physics, Harvard University, Cambridge, MA 02138, USA}

\author{Nobuyuki Yoshioka}
\email{nyoshioka@ap.t.u-tokyo.ac.jp}
\affiliation{Department of Applied Physics, University of Tokyo, 7-3-1 Hongo, Bunkyo-ku, Tokyo 113-8656, Japan}
\affiliation{Quantum Computing Center, RIKEN Cluster for Pioneering Research (CPR), Wako-shi, Saitama 351-0198, Japan}
\affiliation{JST, PRESTO, 4-1-8 Honcho, Kawaguchi, Saitama, 332-0012, Japan}

\begin{abstract}
    We have established the method of characterizing the unitary design generated by a symmetric local random circuit. 
    Concretely, we have shown that the necessary and sufficient condition for the circuit asymptotically forming a $t$-design is given by simple integer optimization for general symmetry and locality. 
    By using the result, we explicitly give the maximal order of unitary design under the $\mathbb{Z}_2$, U(1), and SU(2) symmetries for general locality. 
    This work reveals the relation between the fundamental notions of symmetry and locality in terms of randomness.
\end{abstract}
\maketitle

\let\oldaddcontentsline\addcontentsline
\renewcommand{\addcontentsline}[3]{}

\textit{Introduction}.---
Symmetry is one of the most foundational and ubiquitous principles that affect natural phenomena.
It not only puts constraints on static or dynamic properties of systems on microscopic and macroscopic scales, but also gives rise to new phenomena and techniques such as spontaneous symmetry breaking~\cite{nambu1960axial, nambu1961dynamical, goldstone1961field}, 
symmetry-enhanced topological phenomena~\cite{wen1995topological, kane2005quantum, hasan2010colloquium, sato2017topological, gu209tensor, pollman2010entanglement, chen2012science}, 
quantum memories~\cite{shor1995scheme, steane1996error, calderbank1996good, nielsen2000quantum}, 
decoherence-free subspaces~\cite{zanardi1997noiseless, duan1997preserving, lidar1998decoherence, duan1998reducing, lidar1999concatenating, bacon1999robustness, bacon2000universal, lidar2001adecoherence, lidar2001bdecoherence, kempe2001theory, kempe2001encoded}, covariant quantum error-correcting codes~\cite{woods2020continuous, faist2020continuous, hayden2021error, zhou2021new, kong2022nearoptimal, yang2022optimal, dai2023approximate, liu2023approximate, denys2024quantum}, and
geometric quantum machine learning~\cite{larocca2022group, meyer2023exploiting, sauvage2024building, zheng2023speeding, nguyen2024theory, wang2022symmetric, schatzki2024theoretical, west2024provably, le2023symmetry, wierichs2023symmetric}. 
The concept has an intriguing interplay with yet another fundamental notion, locality, in both spatial and temporal ways. 
The former emerges for instance in the hydrodynamic transport~\cite{rakovszky2018diffusive, agarwal2023charge} and quantum chaos~\cite{roberts2017chaos, cotler2017chaos}, whereas the latter includes the universality of unitary evolution, i.e., the ability to generate arbitrary unitary operations. 
This is a crucial concept in quantum information science including quantum computing, quantum control, and quantum cryptography.

In the absence of any symmetry, 
it is well understood that the universality regarding the global unitary group is achieved by layers of locally universal gate sets~\cite{divincenzo1995universal, lloyd1995almost, barenco1995elementary}. 
While one may naturally deduce that such a connection between globalness and locality is robust, it was shown  
that symmetric local gates are not universal for symmetric global gates~\cite{marvian2022restriction}.
This counterintuitive no-go theorem of universality casts light on the impact of symmetry on the practicality of quantum information processing and quantum computing~\cite{nakata2023black}, as well as the statistical and dynamical properties of quantum many-body systems~\cite{lavasani2021measurement, sang2021measurement, guryanova2016thermodynamics, yungerhalpern2016microcanonical, mitsuhashi2022characterizing, shayan2023critical, varikuti2024unraveling}.
An outstanding question we must ask is the following: {\it to what extent the globalness of unitary evolution is allowed under the presence of symmetry for individual quantum operations?}

A first key step was made by introducing a weaker version of the universality called {\it semi-universality}, which means the relaxed tunability up to the relative phases between symmetry sectors~\cite{kempe2001encoded}.
Such a property is indeed shown to be satisfied in some non-universal cases including 2-local circuits obeying symmetries such as $\mathbb{Z}_2$, U(1) or SU(2) with $2$-local interaction~\cite{marvian2023theory, marvian2024rotationally} and $\mathrm{SU}(d)$ with $3$-local gates~\cite{hulse2024framework}.

A complementary approach to such a qualitative picture is to examine the expressibility in a quantitative manner via the randomness of local circuits. As known in the non-symmetric cases, this allows us to assess the applicability to quantum tomography~\cite{huang2020predicting}, quantum device benchmarking~\cite{emerson2005scalable, knill2008randomized, magesan2011scalable}, variational quantum computation~\cite{mcclean2018barren}, and so on.
In this direction, Refs.~\cite{li2023designs, hearth2025unitary} lower-bounded the expressibility of symmetric circuits with specific locality via the ability to compose a symmetric unitary $t$-design, or, mimic the symmetric Haar random unitary distribution up to the $t$th moment~\cite{emerson2005scalable, mitsuhashi2023clifford, tsubouchi2024symmetric}.
However, in sharp contrast to the non-symmetric case, it remains an open question to establish a unified theory that provides the maximal order of unitary designs achievable with symmetric local random circuits for general symmetry and locality.

In this work, we fill this void by establishing a formula to characterize the randomness of random circuits under arbitrary symmetry and locality constraints.
Concretely, we prove that finding the tight bound on the achievable order of unitary designs can be reduced to a simple integer optimization problem. 
By using this result, we explicitly obtain the maximal order of unitary designs generated by local random circuits under physically important symmetries $\mathbb{Z}_2$, $\mathrm{U}(1)$, and $\mathrm{SU}(2)$.

\begin{figure*}
    \centering
    \includegraphics[width=155mm]{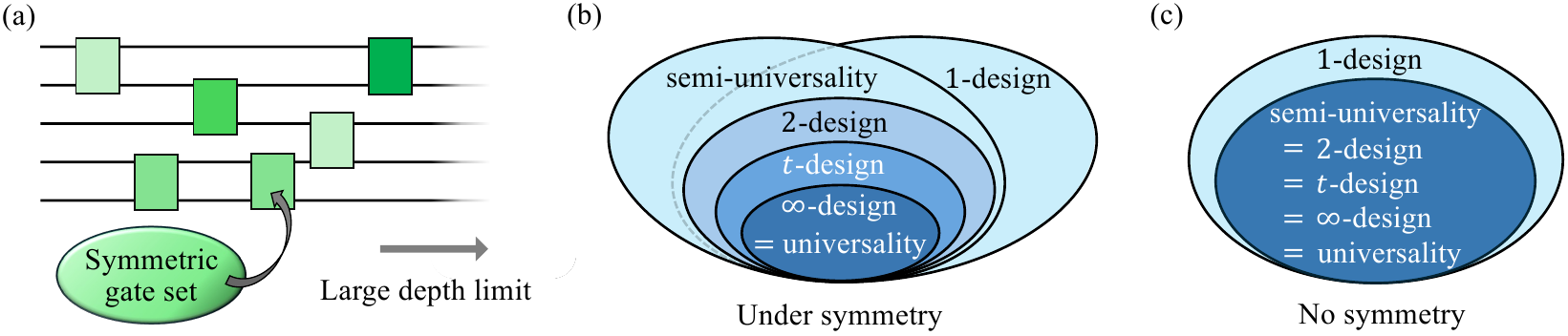}
    \caption{Setup of random circuits and relation between semi-universality and unitary designs with and without symmetry.
    (a) We consider a symmetric local random circuit in the large depth limit. Local random gates are drawn from a symmetric gate set, which is a set of all local gates commuting with a group representing symmetry, such as $\mathbb{Z}_2$, $\mathrm{U}(1)$, and $\mathrm{SU}(2)$.
    (b) In our setup, semi-universality (see Definition~\ref{def:semi_universality} for explanation) and unitary designs have rich relation under symmetry, while (c) the relation becomes rather trivial without symmetry. }
    \label{fig-setting}
\end{figure*}

\textit{Setup}.---
We consider random circuits consisting of symmetric and local gates on $n$ qudits, as shown in Fig.~\ref{fig-setting} (a). 
We first explain the symmetry constraint and next, introduce the locality constraint. 
By using the pair of a group $G$ and its unitary representation $R$, we say that an operator is $(G, R)$-symmetric if it commutes with $R(g)$ for all $g\in G$, and we denote the set of all $(G, R)$-symmetric unitary gates on $n$ qudits by $\mathcal{U}_{n, G, R}$. 
Specifically, the representations of the physically plausible $\mathbb{Z}_2$, $\mathrm{U}(1)$, and $\mathrm{SU}(2)$ symmetries on $n$ qubits are respectively given by 
\begin{align}
    &R(0)=\mathrm{I}^{\otimes n},\ R(1)=\mathrm{Z}^{\otimes n}, \label{eq:Z2_representation}\\
    &R(e^{i\theta})=e^{i\theta \mathrm{Z}^\mathrm{tot}}, \label{eq:U1_representation}\\
    &R(e^{i(\theta_\mathrm{X} \mathrm{X}+\theta_\mathrm{Y} \mathrm{Y}+\theta_\mathrm{Z} \mathrm{Z})})=e^{i(\theta_\mathrm{X} \mathrm{X}^\mathrm{tot}+\theta_\mathrm{Y} \mathrm{Y}^\mathrm{tot}+\theta_\mathrm{Z} \mathrm{Z}^\mathrm{tot})}, \label{eq:SU2_representation}
\end{align} 
where $\mathrm{X}$, $\mathrm{Y}$, and $\mathrm{Z}$ are the Pauli operators, and the superscript ``tot'' means the sum of the Pauli operators acting on every single qubit. 
These symmetries appear in the transverse-field Ising model, the Heisenberg XXZ model, and the XXX model, respectively, for example.

Next, by adding the locality constraints, we define symmetric local random circuits. 
Given $\gamma$ representing a set of labels for qudits, we denote the set of all $(G, R)$-symmetric unitary gates nontrivially acting on the qudit set $\gamma$ by $\mathcal{U}_{n, G, R}^\gamma$. 
We construct a random circuit by taking $\gamma$ from the set $\Gamma$ of qudit sets with nonzero probability $p^\gamma>0$ and randomly drawing a unitary gate $U$ from $\mathcal{U}_{n, G, R}^\gamma$. 
Then, the distribution of unitary gates at each time step is given by 
\begin{align}
    \zeta=\sum_{\gamma\in\Gamma} p^\gamma \mu_{\mathcal{U}_{n, G, R}^\gamma}, \label{eq:local_random_distribution_def}
\end{align}
where $\mu_{\mathcal{U}_{n, G, R}^\gamma}$ is the normalized Haar measure on $\mathcal{U}_{n, G, R}^\gamma$. 
For example, when we express the nearest neighbor gate sets on a $1$-dimensional system consisting of $n$ qudits labeled as $1$, $2$, ..., $n$ with the open boundary condition, $\Gamma$ is given by $\{\{j, j+1\} | j=1, 2, ..., n-1\}$, which corresponds to the circuit shown in Fig.~\ref{fig-setting} (a). 
We define the locality $k$ of $\Gamma$ as the maximal number of qudits that we can simultaneously manipulate, i.e., $k:=\max_{\gamma\in\Gamma} \#\gamma$, where $\#\gamma$ means the size of a set $\gamma$. 
In the following, in order to exclude trivial cases, we assume that the circuit cannot be decomposed into any two independent parts, i.e., if we decompose the qudits into any two parts, there exists some gate set $\mathcal{U}_{n, G, R}^\gamma$ that acts on both parts.

In order to characterize the randomness of the symmetric random circuits specified by the distribution $\zeta$, we rely on a notion called \textit{asymptotic unitary design}, defined as follows.

\begin{definition}
    (Asymptotic symmetric unitary design.) 
    Let $\nu$ be a distribution on the unitary group on $n$ qudits. 
    $\nu$ is an asymptotic $(G, R)$-symmetric unitary $t$-design if 
    \begin{align}
        \lim_{D\to\infty} (M_{t, \nu})^D=M_{t, \mu_{\mathcal{U}_{n, G, R}}}, 
    \end{align}
    where $M_{t, \nu}$ is the $t$th-order moment operator defined by $\int U^{\otimes t}\otimes U^{*\otimes t}d\nu(U)$.
\end{definition}

This condition implies that the behavior of the distribution $\nu$ asymptotically coincides with the symmetric global Haar random distribution up to the $t$th order in the infinite depth. 
Our definition of asymptotic unitary design is equivalent to forming an $\epsilon$-approximate unitary design for all $\epsilon>0$ by taking a sufficiently large depth $D$.
We note that even if we assume the infinite depth, the problem of finding the maximal order of $t$ is still highly nontrivial in the presence of symmetry as opposed to the non-symmetric case, where there does not exist an upper bound on the achievable order $t$~\cite{brandao2016local}. 
\vspace{3mm}

\textit{Results for concrete symmetries}.---
We focus on the $\mathbb{Z}_2$, $\mathrm{U}(1)$, and $\mathrm{SU}(2)$ symmetries in our first main result, which explicitly gives the tight bound on the maximal achievable order of asymptotic unitary design for sufficiently large $n$.

\begin{theorem} \label{thm:special}
    (Concrete symmetries.) 
    Let $n\geq k+1$, $G=\mathbb{Z}_2$, $\mathrm{U}(1)$, or $\mathrm{SU}(2)$, and $R$ be defined by Eq.~\eqref{eq:Z2_representation}, \eqref{eq:U1_representation}, or \eqref{eq:SU2_representation}, respectively. 
    Then, the distributions $\zeta$ defined by Eq.~\eqref{eq:local_random_distribution_def} for $(G, R)$-symmetric $k$-local random circuits form asymptotic $(G, R)$-symmetric unitary $t$-designs if and only if 
    \begin{align}
        &(\mathbb{Z}_2) && t<2^{n-1}, \label{eq:Z2_design_bound}\\
        &(\mathrm{U(1)}) && t<\frac{2^{\lfloor k/2\rfloor}}{\lceil k/2\rceil!}\prod_{\alpha=1}^{\lceil k/2\rceil} (n-k+2\alpha-1), \label{eq:U1_design_bound}\\
        &(\mathrm{SU(2)}) && t<\frac{2^{\lfloor k/2\rfloor}}{(\lfloor k/2\rfloor+1)!}\prod_{\alpha=1}^{\lfloor k/2\rfloor+1} (n-2\alpha+1), \label{eq:SU2_design_bound}
    \end{align}
    where we assume that $n\geq 2^k$ in the $\mathrm{U}(1)$ case and $n\geq 2^{2(\lfloor k/2\rfloor+1)}$ in the $\mathrm{SU}(2)$ case. 
\end{theorem}

We note that this result depends only on the locality $k$ of $\Gamma$, and not on other information about $\Gamma$ such as the connectivity between qubits, or the detailed probability $p^\gamma$, because every symmetric $k$-local unitary can be written as the product of some unitary $U\in\mathcal{U}_{n, G, R}^\gamma$ satisfying $\#\gamma=k$ and some permutation operator, which are also generated by products of swap operators included in some $\mathcal{U}_{n, G, R}^\gamma$. 
In Table~\ref{table:design_bound}, for the cases of $k=2$, $3$, and $4$, we present the ranges of the qubit count $n$ for which Eqs.~\eqref{eq:U1_design_bound} and \eqref{eq:SU2_design_bound} give the tight upper bounds on the achievable orders of unitary designs (see Theorems~4 and 6 of Ref.~\cite{mitsuhashi2025characterization} for smaller qubit count $n$). 
Although Eq.~\eqref{eq:U1_design_bound} gives the tight bound for all $n\geq k+1$ when $k=2$, $3$ and $4$, it is not necessarily tight when $k\geq 5$. 
We finally note that even when $n$ is not sufficiently large and Eqs.~\eqref{eq:U1_design_bound} and \eqref{eq:SU2_design_bound} do not give the tight bound, they give upper bounds on the achievable orders of unitary designs for any $n\geq k+1$.

\begin{table*}[t]
    \caption{Necessary and sufficient conditions on $t$ for the distribution $\zeta$ of the symmetric local random circuit forming an asymptotic symmetric unitary $t$-design.}
    \label{table:design_bound}
    \centering
    \begin{tabular}{l||ll|ll|ll}
         & $2$-local && $3$-local  && $4$-local &\\
        \hline\hline
        $\mathbb{Z}_2$ symmetry   & $t<2^{n-1}$ & ($n\geq 3$) & $t<2^{n-1}$ & ($n\geq 4$)& $t<2^{n-1}$ & ($n\geq 5$)\\
        $\mathrm{U}(1)$ symmetry  & $t<2(n-1)$ & ($n\geq 3$)& $t<n(n-2)$ & ($n\geq 4$)& $t<2(n-1)(n-3)$& ($n\geq 5$)\\
        $\mathrm{SU}(2)$ symmetry   & $t<(n-1)(n-3)$ & $(n\geq 9)$ & $t<(n-1)(n-3)$ & $(n\geq 9)$ & $t<2(n-1)(n-3)(n-5)/3$ & $(n\geq 13)$ \\
    \end{tabular}
\end{table*}

\vspace{3mm}
\textit{Result for general case}.---
We present the result for general symmetries. 
For stating the result, we prepare two notions: semi-universality and symmetric operator decomposition. 
While a unitary gate set is called universal for $\mathcal{U}_{n, G, R}$ if it can generate $\mathcal{U}_{n, G, R}$, the semi-universality is a more relaxed condition defined as follows:

\begin{definition} \label{def:semi_universality}
    (Semi-universality.)
    The gate set $\mathcal{X}\subset\mathcal{U}_{n, G, R}$ is semi-universal for $\mathcal{U}_{n, G, R}$ if 
    \begin{align}
        \braket{\mathcal{X}}\cdot Z(\mathcal{U}_{n, G, R})=\mathcal{U}_{n, G, R}, 
    \end{align}
    where $\braket{\mathcal{X}}$ is the group generated by $\mathcal{X}$, and $Z(\mathcal{U}_{n, G, R})$ is the center of $\mathcal{U}_{n, G, R}$, i.e., the set of all $U\in\mathcal{U}_{n, G, R}$ commuting with every element in $\mathcal{U}_{n, G, R}$. 
\end{definition}

This definition means that a semi-universal gate set can generate all operators in $\mathcal{U}_{n, G, R}$ up to the freedom of $Z(\mathcal{U}_{n, G, R})$, which is the relative phase group between symmetry sectors explicitly shown below as Eq.~\eqref{eq:relative_phase}. 
In the absence of symmetry, i.e., $G=\{I\}$, the condition of semi-universality reduces to the ordinary universality up to the global phase, because the center of unitary group $\mathcal{U}_n$ is the global phase group $\{e^{i\theta}I\}_{\theta\in\mathbb{R}}$. 
The semi-universality was introduced in the context of decoherence-free subspaces~\cite{kempe2001encoded}, and it is known that the $\mathbb{Z}_2$, $\mathrm{U}(1)$, or $\mathrm{SU}(2)$-symmetric $2$-local gate sets are not universal but semi-universal~\cite{marvian2023theory, marvian2024rotationally}.

We clarify the relation between (semi-)universality and unitary design shown in Figs.~\ref{fig-setting} (b) and (c). 
First, if the distribution $\zeta$ defined by Eq.~\eqref{eq:local_random_distribution_def} is an asymptotic $(G, R)$-symmetric unitary $2$-design, the gate set $\bigcup_{\gamma\in\Gamma} \mathcal{U}_{n, G, R}^\gamma$ is semi-universal for $\mathcal{U}_{n, G, R}$, which follows from Theorem~16 of Ref.~\cite{zeier2015squares} (see below Definition~2 in Ref.~\cite{mitsuhashi2025characterization} for details).
The converse is not true; 
As a counterexample, under the $\mathbb{Z}_2$ symmetry on a single qubit, the trivial gate set $\{I\}$ is semi-universal for $\mathcal{U}_{1, G, R}$, but does not generate even a $1$-design.  
Second, the universality of $\bigcup_{\gamma\in\Gamma} \mathcal{U}_{n, G, R}^\gamma$ for $\mathcal{U}_{n, G, R}$ is equivalent to $\zeta$ forming an asymptotic unitary $t$-design for all $t\in\mathbb{N}$, as we explain below Theorem~\ref{thm:general}. 
This implies that in the non-symmetric case, the semi-universality and the universality coincide, and thus all the classes of unitary $t$-designs with $t\geq 2$ collapse into $\infty$-designs. 
We note that this statement holds only for continuous gate sets, and not for discrete gate sets such as the Clifford gates.

Next, we introduce the decomposition of symmetric operators. 
Let us start from the fact that every unitary representation $R$ can be decomposed into inequivalent irreducible representations: 
\begin{align}
    R(g)\cong \bigoplus_{\lambda\in\Lambda} R_\lambda(g)\otimes \mathrm{id}(\mathbb{C}^{m_\lambda})\ \forall g\in G, 
\end{align}
where $\Lambda$ is the set of all labels $\lambda$ for the irreducible representations $R_\lambda$'s that appear in the decomposition of $R$ with multiplicity $m_\lambda$, $\mathrm{id}(\mathbb{C}^m)$ is the identity operator on $\mathbb{C}^m$, and $\cong$ means isomorphism. 
Then, by Schur's lemma, every $(G, R)$-symmetric operator $A$ can be written as~\cite{bartlett2007reference} 
\begin{align}
    A\cong \bigoplus_{\lambda\in\Lambda} \mathrm{id}(\mathbb{C}^{r_\lambda})\otimes A_\lambda \label{eq:sym_op_decomp}
\end{align}
with some operators $A_\lambda$ acting on the multiplicity spaces $\mathbb{C}^{m_\lambda}$, where $r_\lambda$ is the dimension of the representation space of $R_\lambda$, and $A_\lambda$ is uniquely determined for $A$. 
By using this decomposition, we can confirm that $Z(\mathcal{U}_{n, G, R})$ is the relative phase group: 
\begin{align}
    Z(\mathcal{U}_{n, G, R})\cong\left\{\bigoplus_{\lambda\in\Lambda} e^{i\theta_\lambda}\mathrm{id}(\mathbb{C}^{r_\lambda}\otimes\mathbb{C}^{m_\lambda})\ \middle|\ \theta_\lambda\in\mathbb{R}\right\}. \label{eq:relative_phase}
\end{align}

We illustrate the decomposition by taking the $\mathrm{U}(1)$ symmetry. 
Since $\mathrm{U}(1)$ is commutative, all irreducible representations are $1$-dimensional and correspond to the eigenvalues of $\mathrm{Z}^\mathrm{tot}$. 
Concretely, it is decomposed into $n+1$ irreducible representations $R_\lambda(e^{i\theta})=e^{i(n-2\lambda)\theta}$ with $\lambda\in\{0, 1, ..., n\}$, and its multiplicity $m_\lambda$ is given by $\binom{n}{\lambda}$, which corresponds to the dimension of the eigenspace of $\mathrm{Z}^\mathrm{tot}$ with eigenvalue $n-2\lambda$.

We are now ready to state our general theorem.

\begin{theorem} \label{thm:general}
    (General symmetry.)
    Let $\bigcup_{\gamma\in\Gamma} \mathcal{U}_{n, G, R}^\gamma$ be semi-universal for $\mathcal{U}_{n, G, R}$. 
    Then, the distribution $\zeta$ defined by Eq.~\eqref{eq:local_random_distribution_def} is an asymptotic $(G, R)$-symmetric unitary $t$-design if and only if  
    \begin{align}
        t<\min_{\bm{x}\in(\mathcal{C}^\perp\cap\mathbb{Z}^\Lambda)\backslash\{\bm{0}\}} \braket{\bm{m}, \bm{x}^+}, \label{eq:order_bound_general}
    \end{align}
    where $\mathcal{C}:=\mathrm{span}(\{\bm{f}(A) | A\in\mathcal{L}_{n, G, R}^\gamma, \gamma\in\Gamma\})$, $\mathcal{L}_{n, G, R}^\gamma$ is the set of all $(G, R)$-symmetric linear operators nontrivially acting on the qudits labeled by $\gamma$, $\bm{f}(A)=(f_\lambda(A))_{\lambda\in\Lambda}:=(\mathrm{tr}(A_\lambda))_{\lambda\in\Lambda}$ with $A_\lambda$ in Eq.~\eqref{eq:sym_op_decomp} for symmetric operators $A$, $\bm{m}:=(m_\lambda)_{\lambda\in\Lambda}$, $\bm{x}^+=(x_\lambda^+)_{\lambda\in\Lambda}:=((|x_\lambda|+x_\lambda)/2)_{\lambda\in\Lambda}$, we use the standard inner product $\braket{\bm{a}, \bm{b}}:=\sum_{\lambda\in\Lambda} a_\lambda^* b_\lambda$ for $\bm{a}, \bm{b}\in\mathbb{C}^\Lambda$, and $\mathcal{C}^\perp$ is the orthogonal complement of $\mathcal{C}$ with respect to the inner product. 
\end{theorem}

This theorem applies to arbitrary symmetries including noncommutative ones as long as the semi-universality condition holds. 
We note that when the representation $R$ is given by $T^{\otimes n}$ with a single-qudit representation $T$ of $G$, for example in Eqs.~\eqref{eq:Z2_representation}, \eqref{eq:U1_representation}, and \eqref{eq:SU2_representation}, $\mathcal{C}$ can be simplified to $\{\bm{f}(A\otimes \mathrm{I}^{\otimes n-k}) | A\in\mathcal{L}_{k, G, T^{\otimes k}}\}$, where $\mathcal{L}_{k, G, T^{\otimes k}}$ is the set of all linear operators on $k$ qubits commuting with $T^{\otimes k}(g)$ for all $g\in G$. 
In this case, the dimension of $\mathcal{C}$ is constant with respect to the qudit count $n$, because it is no greater than the dimension of $\mathcal{L}_{k, G, T^{\otimes k}}$.

Two remarks are in order. 
First, by Theorem~\ref{thm:general}, we can confirm that $\zeta$ is an asymptotic unitary $t$-design for all $t\in\mathbb{N}$ if and only if $\bigcup_{\gamma\in\Gamma} \mathcal{U}_{n, G, R}^\gamma$ is universal for $\mathcal{U}_{n, G, R}$. 
Concretely, when the gate set is universal, it follows that $\mathcal{C}^\perp=\{\bm{0}\}$, and thus Eq.~\eqref{eq:order_bound_general} means $t<\infty$, where we define the minimization over the empty set as $\infty$. 
On the other hand, when the gate set is not universal, we can construct an upper bound $t_0$ on the r.h.s. of Eq.~\eqref{eq:order_bound_general} as follows: 
We take some $\bm{d}\in(\mathcal{C}^\perp\cap\mathbb{Z}^\Lambda)\backslash\{\bm{0}\}$, and by using $\bm{d}^+:=((|d_\lambda|+d_\lambda)/2)_{\lambda\in\Lambda}$, define $t_0:=\braket{\bm{m}, \bm{d}^+}$, which is an upper bound (see Lemma~14 in Ref.~\cite{mitsuhashi2025characterization} for the proof of the existence of such $\bm{d}$).
Then, $\zeta$ is not an asymptotic unitary $t_0$-design.

Second, the r.h.s. of Eq.~\eqref{eq:order_bound_general} can be calculated by an enumerative method by using the upper bound $t_0$ mentioned above. 
Concretely, we take every $\bm{x}\in (\mathcal{C}^\perp\cap\mathbb{Z}^\Lambda)\backslash\{\bm{0}\}$ satisfying $|x_\lambda|<t_0/m_\lambda$ for all $\lambda\in\Lambda$, and check whether there exists $\bm{x}$ such that $\braket{\bm{m}, \bm{x}^+}<t_0$ by enumeration. 
If there exists such $\bm{x}$, the r.h.s. of Eq.~\eqref{eq:order_bound_general} is given by the minimum of $\braket{\bm{m}, \bm{x}^+}$, and otherwise, it is given by $t_0$. 
This method works because for any $\bm{x}$ satisfying $|x_\lambda|\geq t_0/m_\lambda$ with some $\lambda\in\Lambda$, we have $\braket{\bm{m}, \bm{x}^+}=\max\{\braket{\bm{m}, (\pm\bm{x})^+}\}\geq \max\{m_\lambda (\pm x_\lambda)^+\}= m_\lambda |x_\lambda|\geq t_0$. 
We note that there are some cases when the enumeration is unnecessary, such as $\mathrm{U}(1)$-symmetric case for sufficiently large $n$, which is explained in the proof below.

We find that it is instructive to present the intuitive proof idea of Theorem~\ref{thm:general} (see Theorem~1 of Ref.~\cite{mitsuhashi2025characterization} for the complete proof). 
The notion of asymptotic unitary design is directly connected to the expressibility of the available gate sets. 
Since we assume semi-universality, the difference in the expressibility of symmetric local/global unitaries appears only in terms of the relative phases. 
Therefore, the condition of asymptotic $t$-design reduces to the possibility of estimating the component of the relative phases for any sum of $t$ relative phases. 
This condition is equivalently rephrased as the condition for a certain set of equations having no nontrivial integer solution, which can also be expressed as integer optimization.

\textit{Derivation of Theorem~\ref{thm:special} from Theorem~\ref{thm:general}}.---
Although we guide readers to Theorems~2, 3, and 5 in Ref.~\cite{mitsuhashi2025characterization} for rigorous proofs, we explain the general procedure to get explicit values of the r.h.s. of Eq.~\eqref{eq:order_bound_general} under appropriate conditions on $n$ in three steps. 
Note that the latter two steps are in similar to the numerical method mentioned above.

In the first step, we get the explicit expressions of $\bm{m}$ and the basis of $\mathcal{C}$ by using the representation theory. 
We focus on the case when the size of $\Lambda$ is larger than the dimension $J$ of $\mathcal{C}$, since Eq.~\eqref{eq:order_bound_general} trivially means $t<\infty$ otherwise. 
In the second step, we take an upper bound $t_0$ on r.h.s. of Eq.~\eqref{eq:order_bound_general} as follows: 
we define $\Lambda'$ as the set of labels $\lambda$ corresponding to $J+1$ smallest $m_\lambda$'s, and define the $(J+1)$-dimensional space $\mathcal{D}:=\{\bm{x}\in\mathbb{R}^\Lambda | x_\lambda=0\ \forall \lambda\in\Lambda\backslash\Lambda'\}$. 
Then, $\mathcal{C}^\perp\cap\mathbb{Z}^\Lambda\cap\mathcal{D}$ is a $1$-dimensional lattice. 
We take a nonzero lattice point $\bm{d}$ closest to $\bm{0}$, and define $t_0:=\braket{\bm{m}, \bm{d}^+}$. 
In the third step, we show that the r.h.s. is exactly given by $t_0$ if $m_\lambda\geq t_0$ for all $\lambda\in\Lambda\backslash\Lambda'$, which holds for sufficiently large $n$ in the $\mathrm{U}(1)$ and $\mathrm{SU}(2)$ cases. 
This can be confirmed by noting that if we take arbitrary $\bm{x}\in(\mathcal{C}^\perp\cap\mathbb{Z}^\Lambda)\backslash\mathcal{D}$, then we have $\braket{\bm{m}, \bm{x}^+}\geq m_\lambda |x_\lambda|\geq t_0$.

We illustrate this procedure by taking the $\mathrm{U}(1)$-symmetric and $2$-local case as the simplest nontrivial example. 
First, as we have mentioned just above Theorem~\ref{thm:general}, we have $\Lambda=\{0, 1, ..., n\}$, and the multiplicity $m_\lambda$ is given by $\binom{n}{\lambda}$. 
In order to take the basis of $\mathcal{C}$, we note that every $A\in\mathcal{L}_{2, G, T^{\otimes 2}}$ can be written in a block diagonal form $A_0\oplus A_1\oplus A_2$ with $A_j$'s acting on the eigenspaces of $\mathrm{Z}\otimes\mathrm{I}+\mathrm{I}\otimes\mathrm{Z}$ with eigenvalues $2-2j$, respectively. 
Thus, $f_\lambda(A\otimes\mathrm{I}^{\otimes n-2})$ is given by $\sum_{j=0}^2 \mathrm{tr}(A_j)\binom{n-2}{\lambda-j}$, which implies that we can take $\bm{c}_j=(\binom{n-2}{\lambda-j})_{\lambda\in\Lambda}$ with $j=0, 1, 2$ as a basis of $\mathcal{C}$, where we define $\binom{a}{b}:=0$ when $b\not\in\{0, 1, ..., a\}$. 
Second, we take an upper bound on the r.h.s. of Eq.~\eqref{eq:order_bound_general}. 
Since $\mathcal{C}$ is $3$-dimensional, we define $\Lambda':=\{0, 1, n-1, n\}$ with its size $4$. 
Then, the set of $\bm{x}\in\mathcal{C}^\perp\cap\mathbb{Z}^\Lambda$ satisfying $x_\lambda=0$ for all $\lambda\in\Lambda\backslash\Lambda'$ forms a $1$-dimensional lattice $\{a\bm{d} | a\in\mathbb{Z}\}$ with $\bm{d}:=(1, -(n-2), 0, ..., 0, n-2, -1)$. 
We define $t_0:=\braket{\bm{m}, \bm{d}^+}=2(n-1)$. 
Third, we show that it equals the r.h.s. of Eq.~\eqref{eq:order_bound_general} when $n\geq 4$. 
For the proof of this, it is sufficient to confirm that when $\bm{x}$ is not in the $1$-dimensional lattice mentioned above, we can take some $\lambda\in\Lambda\backslash\Lambda'$ satisfying $x_\lambda\neq 0$, and we have $\braket{\bm{m}, \bm{x}^+}\geq m_\lambda |x_\lambda|\geq m_\lambda\geq \binom{n}{2}\geq 2(n-1)=t_0$. 
We can use a similar method for general locality. 
We present the proof for the $\mathbb{Z}_2$ and $\mathrm{SU}(2)$ symmetries in Appendix~\ref{sec:Z2_SU2}.

\textit{Discussion}.---
In this work, we have presented a formula for finding the maximal order of unitary design generated with symmetric local random circuits for general symmetries and general localities in Theorem~\ref{thm:general}. 
By applying this result to the $\mathbb{Z}_2$, $\mathrm{U}(1)$, and $\mathrm{SU}(2)$ symmetries, we have explicitly shown the maximal order of unitary designs for general localities in Theorem~\ref{thm:special}. 
Although we have focused on a continuous gate set $\{\mathcal{U}_{n, G, R}^\gamma\}_{\gamma\in\Gamma}$, our results are also applicable to random circuits with arbitrary gate sets as long as they generate the continuous unitary ensemble $\{\mathcal{U}_{n, G, R}^\gamma\}_{\gamma\in\Gamma}$. 
Since Theorem~\ref{thm:general} is given in a general form, it can be used to derive the results for other symmetries.

Our result can serve as a tool for distinguishing whether we can use some gate set for the symmetric versions of quantum information processing tasks, which may, for example, lead to extending the symmetric classical shadow~\cite{hearth2024efficient, sauvage2024classical} to other symmetries and gate sets. 
We note that symmetric Clifford gates do not necessarily generate symmetric unitary $3$-designs or even $2$-designs~\cite{mitsuhashi2023clifford}, but instead we may be able to use symmetric local random gates to generate symmetric unitary designs. 
Moreover, since our results hold for any order of unitary design, it would also be useful for investigating the symmetric versions of notions involving $4$th or higher-order unitary designs, such as higher-order randomized benchmarking~\cite{nakata2021quantum} and out-of-time-order higher point correlators~\cite{roberts2017chaos}.

As future research directions, it is important to investigate the convergence speed of the asymptotic unitary designs on symmetric local random circuits.
Additionally, it would be interesting to investigate the effect of our result on entanglement growth~\cite{liu2018entanglement} and deep thermalization~\cite{ippoliti2022solvable} under symmetric local random unitary dynamics. 
Clarifying the constraints deriving from the existence of the upper bound on the order of unitary designs will enable us to gain a deeper understanding of the many-body dynamics under Hamiltonians composed of symmetric local interactions.

\textit{Acknowledgements}.---
The authors wish to thank 
Iman Marvian, 
Hiroyasu Tajima, 
Janek Denzler,
and Zongping Gong
for insightful discussions. 
Y.M. is supported by JSPS KAKENHI Grant No. JP23KJ0421.
R.S. is supported by the BMBF (PhoQuant, Grant No. 13N16103).
This research is funded in part by the
Gordon and Betty Moore Foundation's EPiQS Initiative,
Grant GBMF8683 to T.S.
N.Y. wishes to thank JST PRESTO No. JPMJPR2119, JST ASPIRE Grant Number JPMJAP2316,
and the support from IBM Quantum.
This work was supported by JST Grant Number JPMJPF2221, JST ERATO Grant Number JPMJER2302, and JST CREST Grant Number JPMJCR23I4, Japan.

Note Added: During the preparation of this article, we became aware of independent work by 
Austin Hulse, Hanqing Liu, and Iman Marvian ~\cite{hulse2024unitary}, which studies similar questions and was posted on arXiv concurrently with the present paper.
Both have arrived at the same result on the maximal order of unitary designs under the $\mathrm{U}(1)$ and $\mathrm{SU}(2)$ symmetries. 
Reference~\cite{hulse2024unitary} has assumed conjectures about combinatorial identities, which are introduced as Eqs.~(86) and (120) of the version~1 of their manuscript for the proof of general $k$-local cases.
In our work, we have provided a proof that is independent of any conjectures.

\bibliography{bib.bib}

\appendix

\section{Application of Theorem~\ref{thm:general} to concrete symmetries} \label{sec:Z2_SU2}

In this appendix, we explain how we can apply the result of Theorem~\ref{thm:general} to get Theorem~\ref{thm:special} in the $\mathbb{Z}_2$ and $\mathrm{SU}(2)$ symmetries. 
First, as for the $\mathbb{Z}_2$ symmetry, since $\mathbb{Z}_2$ is commutative, every irreducible representation is one-dimensional, and corresponds to an eigenvalue of $\mathrm{Z}^{\otimes n}$. 
Thus $R$ defined by Eq.~\eqref{eq:Z2_representation} is decomposed into two irreducible representations with $\lambda\in\{0, 1\}$ defined by $R_\lambda(g)=(-1)^{\lambda g}$ for $g\in\{0, 1\}$, and the multiplicity $m_\lambda$ is given by $2^{n-1}$. 
We note that every $A\in\mathcal{L}_{k, G, T^{\otimes k}}$ can be written as $A=A_0\oplus A_1$ with $A_j$'s acting on the eigenspaces of $\mathrm{Z}^{\otimes k}$ with eigenvalues $(-1)^j$. 
Since we have $f_0(A)=f_1(A)=2^{n-k-1}(\mathrm{tr}(A_0)+\mathrm{tr}(A_1))$, we find that $\mathcal{C}$ is spanned by $(1, 1)$. 
Thus Eq.~\eqref{eq:order_bound_general} implies that $t<\min_{\bm{x}\in\{(a, -a) | a\in\mathbb{Z}\backslash\{0\}\}} 2^{n-1}(x_0^+ +x_1^+)=2^{n-1}$.

Next, we consider the $\mathrm{SU}(2)$ case. 
Since we can apply the method similar to the one in the $\mathrm{U}(1)$ symmetry in the second and the third steps, we only explain the first step of explicitly rewriting Eq.~\eqref{eq:order_bound_general}. 
By utilizing the $\mathfrak{sl}(2, \mathbb{C})$ representation theory, the representation $R$ defined by Eq.~\eqref{eq:SU2_representation} can be decomposed into the spin-$\lambda$ representations with $\lambda=\{n/2, n/2-1, ..., n/2-\lfloor n/2\rfloor\}$. 
The multiplicity $m_\lambda$ is given by $\binom{n}{n/2-\lambda}-\binom{n}{n/2-\lambda-1}$, which corresponds to the difference between the dimensions of the eigenspaces of $\mathrm{Z}^\mathrm{tot}$ with eigenvalues $2\lambda$ and $2(\lambda+1)$. 
In order to take a basis $\{\bm{c}_j\}$ of $\mathcal{C}$, we note that $\mathcal{L}_{k, G, T^{\otimes k}}$ can be spanned by permutation operators $Q_\sigma$ on $k$ qubits by the Schur-Weyl duality. 
Thus $\mathcal{C}$ can be spanned by $\bm{f}(Q_\sigma\otimes \mathrm{I}^{\otimes n-k})$. 
By using the properties of the binomial coefficients, we can show that $\mathcal{C}$ can be spanned by $\bm{f}([(I-\mathrm{SWAP})/2]^{\otimes j}\otimes \mathrm{I}^{\otimes n-2j})$ with $j=0, 1, ..., \lfloor k/2\rfloor$ (see Lemma~9 in Ref.~\cite{mitsuhashi2025characterization} for details). 
Since $[(I-\mathrm{SWAP})/2]^{\otimes j}$ is the projection onto the space of the spin-$0$ representation on $2j$ spins, $f_\lambda([(I-\mathrm{SWAP})/2]^{\otimes j}\otimes \mathrm{I}^{\otimes n-2j})$ corresponds to the multiplicity of the spin-$\lambda$ representation in the tensor product of representation of $\mathrm{SU}(2)$ on the rest $n-2j$ qubits, which is given by $\binom{n-2j}{(n-2j)/2-\lambda}-\binom{n-2j}{(n-2j)/2-\lambda-1}$. 

\end{document}